\documentclass[aps,pre,twocolumn,superscriptaddress]{revtex4}
\usepackage{amssymb}
\usepackage{amsmath}
\usepackage{times}
\usepackage{color}
\usepackage{graphicx}
\usepackage{makecell}
\usepackage{threeparttable}
\usepackage{multirow}

\begin{document}

\title{An effective method for profiling core-periphery structures in complex networks}
\author{Jiaqi Nie}
\affiliation{Institute of Cyberspace Security, Zhejiang University of Technology, Hangzhou, 310023, China}
\affiliation{Binjiang Institute of Artificial Intelligence, Zhejiang University of Technology, Hangzhou, 310056, China}

\author{Qi Xuan}
\affiliation{Institute of Cyberspace Security, Zhejiang University of Technology, Hangzhou, 310023, China}
\affiliation{Binjiang Institute of Artificial Intelligence, Zhejiang University of Technology, Hangzhou, 310056, China}

\author{Dehong Gao}
\affiliation{School of Cybersecurity, Northwestern Polytechnical University, Xi’an 710072, China}
\affiliation{Binjiang Institute of Artificial Intelligence, Zhejiang University of Technology, Hangzhou, 310056, China}

\author{Zhongyuan Ruan}
\email{zyuan.ruan@gmail.com}
\affiliation{Institute of Cyberspace Security, Zhejiang University of Technology, Hangzhou, 310023, China}
\affiliation{Binjiang Institute of Artificial Intelligence, Zhejiang University of Technology, Hangzhou, 310056, China}

\begin{abstract}
Profiling core-periphery structures in networks has attracted significant attention, leading to the development of various methods. Among these, the rich-core method is distinguished for being entirely parameter-free and scalable to large networks. However, the cores it identifies are not always structurally cohesive, as they may lack high link density. Here, we propose an improved method building upon the rich-core framework. Instead of relying on node degree, our approach incorporates both the node's coreness $k$ and its centrality within the $k$-core. We apply the approach to twelve real-world networks, and find that the cores identified are generally denser compared to those derived from the rich-core method. Additionally, we demonstrate that the proposed method provides a natural way for identifying an exceptionally dense core, i.e., a clique, which often approximates or even matches the maximum clique in many real-world networks. Furthermore, we extend the method to multiplex networks, and show its effectiveness in identifying dense multiplex cores across several well-studied datasets. Our study may offer valuable insights into exploring the meso-scale properties of complex networks.
\end{abstract}

\maketitle

\section{Introduction}

Investigating the meso-scale structure of networks is crucial for understanding their properties. Of the meso-scale features, community structure is one of the most well-known and has drawn significant attention from researchers \cite{Newman2004,Newman2006,Fortunato2022,Fortunato2016,Shang2020}. Community structure refers to nodes that are densely connected within a group, but sparsely connected to nodes in other groups. The core-periphery structure is another important type of meso-scale feature, which suggests that the network can be divided into two parts \cite{Borgatti2000,Rombach2017,Holme2005,Elliott2020,Gallagher2021,Polanco2023}: a densely connected core and a sparsely connected periphery. Unlike community structure, the nodes in the core are often well-connected not only to each other but also reasonably well-connected to the nodes in the periphery \cite{Rombach2017}. This distinctive structure has been observed in numerous real-world networks, including the world airline network, which features a small, almost fully connected core (accounting for approximately $2.5\%$ of the airports) surrounded by an extensive, nearly tree-like periphery \cite{Verma2014,Verma2016}, as well as the world trade network \cite{Fagiolo2010}, the autonomous internet network \cite{Rossa2013}, and others \cite{Elliott2014,Masuda2006}. 

Many approaches have been proposed to profile the core-periphery structure in networks \cite{Borgatti2000,Lee2014,Kojaku2017,Zhang2015,Ma2015,Mondragon2017}. However, most are complex and may face challenges when applied to very large networks. To address this, Ma and Mondragón developed a simple, fast, and parameter-free method, known as the rich-core method, to effectively profile the core-periphery structure \cite{Ma2015}. This method is pragmatic and highly suitable for application to large-scale networks. In their algorithm, nodes are ranked in descending order based on degree, and their links are categorized as connecting to higher or lower-ranked nodes. A core is identified by locating the node ranked at $r^*$, where the number of links to higher-ranked nodes is maximized. The underlying idea is intuitive and can be linked to random walks in the network. Specifically, the persistence probability of a cluster (the likelihood that a random walker remains within the cluster) increases with the cluster's size. This growth, however, transitions from rapid to gradual at a specific point, where the second derivative of the persistence probability equals to zero. This critical point defines the boundary of the rich core.

The above method has been successfully applied to various real-world systems, ranging from collaboration networks to biological networks \cite{Ma2015pnas,Battiston2018}. Despite its high efficiency in identifying core-periphery structures in networks, the identified cores may not always be structurally cohesive. In this paper, we propose an improved method that retains many advantages of the rich-core approach, such as its non-parametric nature and suitability for large-scale networks, while achieving a denser core. In the following, we first apply our method to a number of real-world single-layer networks and demonstrate its superiority in identifying dense cores. Beyond this, we find that our method naturally provides an effective way to detect exceptionally dense cores, i.e., cliques, within networks. Remarkably, the identified cliques are often very close to, or even equal to the maximum cliques in many networks. We then extend our method to multiplex networks and demonstrate its effectiveness in identifying multiplex cores in the complex systems.


\section{Method}
Consider an unweighted and undirected graph $G=(V,E)$, where $V$ denotes the set of nodes and $E$ the set of links. For each node $i$ in the network, its coreness $k_i$ is calculated using the $k$-core decomposition algorithm \cite{Kitsak2010}. Next, we analyze how a node with coreness $k$ ($k=1,2,...$) interacts with other nodes within the $k$-core --- the subgraph consisting of nodes whose coreness is at least $k$. To do so, we employ various node centrality measures within the $k$-core, denoted as $\widetilde{m}$. In general, the choice of centrality metrics can influence core identification outcomes, as different networks may benefit from different centrality measures. Here, we use degree and eigenvector centrality as examples. As we will see, they both exhibit strong performance in identifying dense cores on the considered datasets. Based on this information, we define the richness (or importance) of node $i$ as $\boldsymbol{\mu}_i=(k_i,\widetilde{m}_i)$. Then, we rank the nodes in the network in descending order of their richness. Specifically,  nodes are first ranked by their coreness values $k_i$, and for nodes with the same coreness, we use their centrality measure $\widetilde{m}_i$ to further rank them. The rank of node $i$ is represented as an integer $r_i$, where nodes with smaller $r_i$ indicate greater richness. For example, the top-ranked node has the highest richness, and the second-ranked node has the next highest richness, and so on. It is worth noting that, multiple nodes may share the same value of richness. In such cases, these nodes are ranked randomly, which may introduce variations across different realizations. However, these differences are inherent to node ranking methods and generally unimportant. Notably, using continuous centrality measures (such as eigenvector centrality) can mitigate this issue.

For each node $i$, its links are divided into two groups: those connecting to nodes with a higher rank (denoted as $d_i^+$), and those connecting to nodes with a lower rank (denoted as $d_i^-$). Consequently, the degree of node $i$ can be expressed as $d_i=d_i^++d_i^-$. Starting with the first node ($r=1$), we sequentially calculate $d_i^+$ for each subsequent node and plot it as a function of its rank, $r_i$. From this analysis, a turning point $r^*$ can be identified, corresponding to the maximal value of $d^+$. This turning point determines the core-periphery boundary of the network: all nodes with ranks smaller than $r^*$ are assigned to the core, while the remaining nodes (with ranks greater than $r^*$) are assigned to the periphery. 

\begin{figure}
\includegraphics[width=1.0\linewidth]{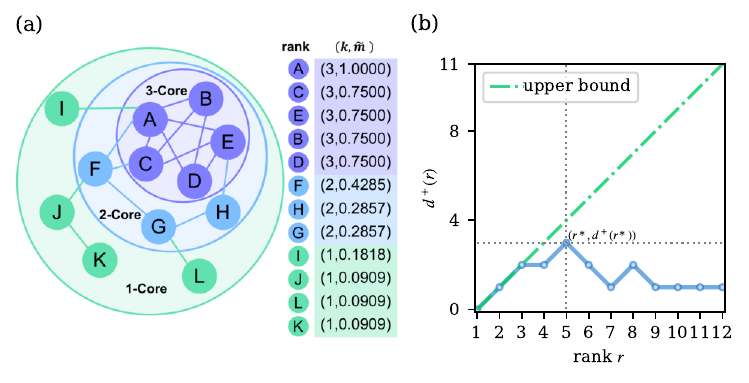} \caption{(a) An illustrative example for ranking nodes based on their richness. Here the richness of a node is defined as a combination of the node coreness $k$ and its degree centrality within the $k$-core, denoted as $\widetilde{m}$. (b) Number of links $d^+$ that a node ranked at $r$ connects to nodes with higher ranks as a function of node rank. The value of $r$ corresponding to the maximum $d^+(r)$ (denoted as $r^*$) defines the boundary of the core-periphery structure. The dot-dash line represents $d^+(r) = r - 1$, which is the upper limit of the $d^+(r)$ curve. Nodes ranked at the top that fall exactly on this line collectively form a clique.} \label{Fig:1}
\end{figure}

As an illustrative example, Figure \ref{Fig:1} (a) demonstrates the detailed ranking process for a small graph consisting of $12$ nodes. For instance, consider node $F$, which has a coreness value of $2$, indicating that it belongs to the $2$-core (comprising nodes $A$ to $H$). The degree centrality of this node within the $2$-core can be calculated as $\widetilde{m}=\frac{\widetilde{d}}{\widetilde{n}-1}=3/7$, where $\widetilde{d}$ is the degree of the node within the $2$-core, and $\widetilde{n}$ is total number of nodes in the $2$-core. Therefore, the richness of node $F$ is represented as $\boldsymbol{\mu}_F=(2,3/7)$. Following this approach, we compute the richness of all nodes and rank them in descending order. We then calculate $d^+$ for each node sequentially and plot it as a function of node rank, as shown in Fig. \ref{Fig:1} (b). The maximum value of $d^+$ occurs at $r=5$, which defines the boundary of a core. 

Furthermore, this method allows us to effectively identify an exceptionally dense core, specifically a clique, within the network. Note that the upper bound of $d_i^+(r_i)$ for any node $i$ (ranked at $r_i$) is $r_i-1$, as it connects to all nodes ranked before it. This implies that the $d^+(r)$ curve always lies below the line $d^+(r) = r-1$. The nodes ranked at the front (starting with the node ranked first) that lie exactly on the line $d^+(r) = r-1$ collectively form a clique. For example, as illustrated in Fig. \ref{Fig:1} (b), the first three nodes together constitute a clique. This initially identified clique can be further expanded through an additional process: we examine subsequent nodes one by one to determine whether each newly considered node connects to all nodes in the existing clique. If it does, the node is added to the clique; otherwise, the process continues with the next node until the end of the rank. As we will show below, the identified clique is very close to (or even equal to) the maximum clique in many real-world networks.

\section{Experiments on single-layer networks}
\subsection{Datasets}
In this section, we will apply our method to various real-world networks. Specifically, we consider the following 12 datasets:\\
1.~{\bf Karate Club} \cite{Zachary1977} . This is a widely studied social network in network science. Nodes in the network represent the members of the karate club, and links denote social ties or friendships between pairs of members.\\
2.~{\bf Dolphins} \cite{Doph2003}. This dataset describes the social interactions within a group of dolphins, illustrating their behavioral connections. \\
3.~{\bf Les Misérables} \cite{Knuth1993} . This is a social network which describes the co-appearance of characters in Les Misérables, the novel by Victor Hugo.\\
4.~{\bf Jazz} \cite{Gleiser2003}. This dataset portrays relationships among jazz musicians who belong to the same band. \\
5.~{\bf Facebook} \cite{Viswanath2009}. This dataset consists of ``circles" (or ``friend lists") for a subset of users on Facebook. \\
6.~{\bf Email} \cite{Guimera2003}. This is an email communication network at Rovira i Virgili University in southern Catalonia, Spain. \\
7.~{\bf Netscience} \cite{newman2006finding}. This is a co-author network composed of scientists engaged in network theory and experiments. Nodes represent researchers, and a link between two nodes indicates that the corresponding researchers have co-authored at least one paper together.\\
8.~{\bf CoAuthorsDBLP} \cite{CiteSeer}. This is a co-author network extracted from the DBLP (Digital Bibliography \& Library Project) computer science bibliography.\\
9.~{\bf CoAuthorsCiteseer} \cite{CiteSeer}. This is a co-author network derived from the Citeseer database.\\
10.~{\bf IPv4}  \cite{CAIDA}. This is a communication network representing the structure of IPv4 routing. Nodes represent Autonomous Systems (AS), and edges represent the connectivity between them.\\
11.~{\bf In-2004} \cite{KONECT}. This is a hyperlink network of the .in domain in India. Nodes represent individual web pages using the .in domain, and edges represent hyperlinks between them. \\
12.~{\bf Wiki-Talk} \cite{KONECT}. This is a communication network of the English Wikipedia. Nodes represent users, and an edge indicates that one user has posted a message on another user's talk page.  \\

The basic properties of these networks are summarized in Table I, including the number of nodes $V$, the number of links $E$, the maximum node coreness $M_{core}$, and the size of the maximum clique $S_{mc}$ in each network. The method we use to determine the maximum clique is based on the Bron-Kerbosch algorithm and its improved version \cite{Bron1973,Tomita2006,Cazals2008}. In the worst-case scenario, the time complexity of the method is $O(3^{n/3})$, where $n$ is the number of nodes in the network, making it computationally expensive for large-scale networks in some cases.

\begin{table*}[ht]
\centering
\caption{The properties of the networks, including the number of nodes ($V$), the number of links ($E$), the maximum node coreness ($M_{core}$), and the size of the maximum clique ($S_{mc}$). The size of the identified cores ($V_{core}$), the core density coefficient ($\phi$), and the size of the identified cliques ($S_{c}$) under different methods.}

\label{tab:Datasets}
\resizebox{\linewidth}{!}{
\begin{tabular}{c|cccc|ccc|ccc|ccc}
\toprule

\multirow{2}*{Datasets}  & \multirow{2}*{$V$} & \multirow{2}*{$E$}  &\multirow{2}*{\makecell[c]{$M_{core}$}} & \multirow{2}*{\makecell[c]{$S_{mc}$}}  &\multicolumn{3}{c|}{rich-core}&\multicolumn{3}{c|}{MCC-E}&\multicolumn{3}{c}{MCC-D}\\

&&&&& $V_{core}$& $\phi$&$S_{c}$&$V_{core}$& $\phi$&$S_{c}$&$V_{core}$& $\phi$&$S_{c}$\\  \cline{1-14}

Karate Club      & 34   & 78 & 4 & 5   &9&0.5556&3&5&1.0000&\textbf{5}&6&0.8000&\textbf{5}
\\
Dolphins              & 62 & 159 & 4 & 5  &24&0.2210&4&12&0.4849&4&20&0.2631& 4
\\
Les\ Misérables           & 77 & 254 & 9 & 10  & 17&0.5367&4&10&1.0000&\textbf{10} &10&1.0000& \textbf{10} 
\\
Jazz    & 198&	2,742	&29	&30	&82&0.3716&9&30&1.0000&\textbf{30}&30&1.0000&\textbf{30}
\\
Facebook  	&4,039	&88,234	&115	&69	&273 &0.3854&7&140&0.9355&67&139&0.9369&67
\\
Email 	&1,133	&5,451	&11	&12	&154&0.0825&4&184&0.0816&\textbf{12}&136 &0.1045&\textbf{12}
\\
Netscience 	&1,461	&2,743	&19	&20 &27&0.5499&9&20&1.0000&\textbf{20}&20&1.0000&\textbf{20}
\\
coAuthorsDBLP   &299,067 &977,676&114& 115&311&0.1627&26&115&1.0000& \textbf{115}&115&1.0000& \textbf{115}
\\  
coAuthorsCiteseer 	 &227,320&814,134&86&87&489& 0.0515&3&87&1.0000&\textbf{87}&87&1.0000&\textbf{87}
\\
IPv4  	&46,172	&176,994	&76&	44	&210&0.3381&18&133&0.7358&39&104&0.8081&38
\\
In-2004   &1,382,908& 13,591,473&488&489&3,758&0.0406&3&489&1.0000&\textbf{489}&489&1.0000&\textbf{489}
\\
Wiki-Talk   &2,394,385  &4,659,565 &131 &26 &1,490&0.1081&7&628&0.3254&24&298&0.5166&23
\\
\cline{1-14}

\end{tabular}}

\end{table*}

\subsection{Results}
\begin{figure}
\includegraphics[width=1.0\linewidth]{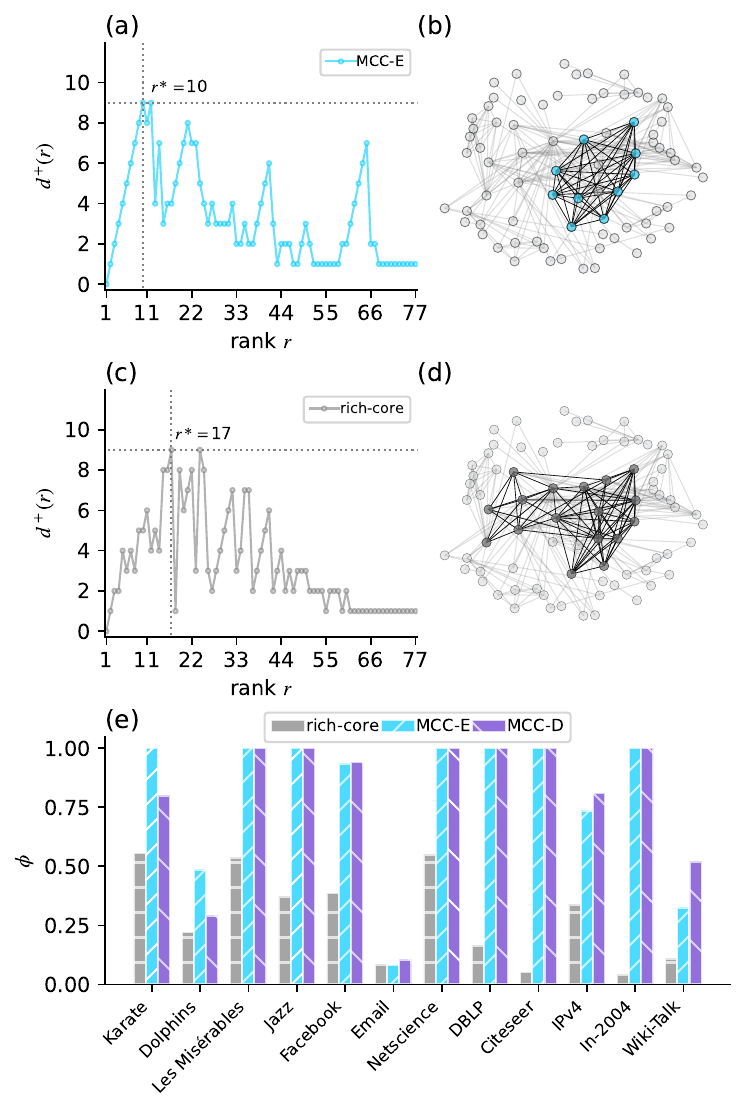} \caption{Number of links $d^+(r)$ that a node ranked at $r$ connects to nodes with higher ranks as a function of node rank for the Les Misérables dataset, under the methods of (a) MCC-E, and (c) rich-core. (b) and (d) present the visualizations of the cores identified by the MCC-E and rich-core methods, respectively. (e) Density of the identified core [as defined in Eq. (\ref{density})] for each network is compared across different methods, including rich-core, MCC-D and MCC-E.} \label{Fig:2}
\end{figure}

For simplicity, hereafter we will refer to the method proposed in section II as MCC, where the richness of a node is determined by its coreness $k$ and its centrality measures within the $k$-core. In our experiments, we specifically focus on degree centrality (denoted as MCC-D) and eigenvector centrality (denoted as MCC-E). Using the dataset of Les Misérables as an example, Figure 2(a) illustrates the curve $d_i^+$ as a function of node rank $r_i$ based on MCC-E. For comparison, Figure 2(c) presents the results obtained using the rich-core method \cite{Ma2015}, where the richness of a node is simply defined by its degree. It is evident that the cores identified by the two methods differ. For instance, the core identified by our method has a size of $r^*=10$, which is smaller than the size of $17$ identified by the rich-core method. To measure the density of the core, we define the density coefficient (also known as the rich-club coefficient) as follows
\begin{eqnarray}\label{density}  
\phi=\frac{2E_{core}}{V_{core}(V_{core}-1)},
\end{eqnarray}
where $V_{core}$ and $E_{core}$ denote the number of nodes and links in the core, respectively. Notice that $\phi \in [0,1]$, and a higher value of $\phi$ indicates a more tightly connected core. Our results show that the core identified by MCC-E is significantly denser ($\phi = 1.0$) compared to the one identified using the rich-core method ($\phi = 0.52$), as illustrated in Fig. \ref{Fig:2}(b) and (d). The superior performance of MCC-E stems from the intrinsic structural properties of $k$-core decomposition: nodes with high coreness values (ranked at top) tend to be well connected to each other, whereas high-degree nodes are not necessarily interconnected.

\begin{figure}
\includegraphics[width=1.0\linewidth]{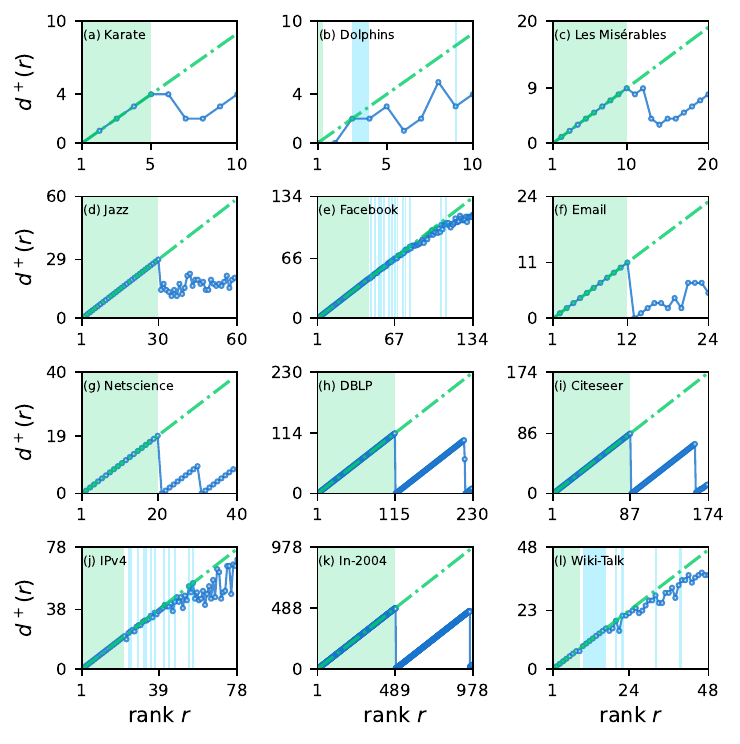} \caption{$d^+(r)$ as a function of node rank $r$ across all $12$ networks under the MCC-E method. The dot-dash line represents $d^+(r) = r - 1$, which is the upper bound of the $d^+(r)$ curve. Nodes within the green shaded region that successively align with the dot-dash line are fully connected, which form a clique. This initial clique can be further expanded by examining subsequent nodes in the rank to see if they connect to all nodes in the existing clique. Nodes that meet this condition are highlighted in the light blue shaded areas.} \label{Fig:3}
\end{figure}

Table I provides detailed information about the identified cores across all $12$ datasets, obtained using the rich-core, MCC-D, and MCC-E methods. Generally, compared to the rich-core method, the cores identified by our method contain fewer nodes, while exhibit a higher density coefficient (meaning that they are denser). A visualization of the density coefficient for these cores is shown in Fig. 2(e). Notice that our method maintains robust performance regardless of network scale, as demonstrated in large networks such as In-2004 and Wiki-Talk. Moreover, it is observed that MCC-D and MCC-E yield the same results ($\phi=1.0$) for half of the datasets. However, in certain cases (e.g., IPv4 networks), MCC-D outperforms MCC-E, whereas in others (e.g., Dolphins), MCC-E shows better performance.

It is important to note that the core identified by our method may contain a denser substructure. Taking the MCC-E method as an example, we plot $d^+(r)$ as a function of node rank $r$ for a subset of high-ranking nodes across all $12$ datasets. The top-ranked nodes that fall precisely on the line $d^+(r) = r-1$ form a clique, as shown in Fig. \ref{Fig:3} (the green shaded areas). Using the additional operation introduced in Sec. II, the clique can be further expanded by incorporating the subsequent nodes (the light blue shaded areas in Fig. \ref{Fig:3}). Finally, the sizes of these cliques (denoted as $S_c$) are listed in Table I. Remarkably, our method proves highly efficient: in $8$ out of the $12$ networks, it successfully identifies the maximum clique. For the remaining networks, the cliques identified are only slightly smaller than the maximum cliques. It should be emphasized that the rich-core method (with the additional operation) is not effective at identifying the cliques. Given that finding the maximum clique in a network is an NP-hard problem, our method offers a simple yet effective approach to approximate this solution. This can be understood by observing that nodes in a clique, especially one of large size, typically have high coreness values and are, therefore, more likely to be ranked near the top.

\section{Experiments on multiplex networks}

The above method can be easily extended to multiplex networks, which are a special case of multilayer networks and are widely used to model various real-world systems \cite{Boccaletti2014,Lee2015,Shen2020,Ruan2015}, such as social networks, infrastructure networks, and biology networks. In a multiplex network, the same set of nodes is connected by multiple layers, with each layer representing a distinct type of link or relationship \cite{Shang2020b,Shang2022}. Let us consider a multiplex network consisting of $M$ layers. The adjacency matrix of each layer $\alpha$ is denoted as $A^{[\alpha]}=\{a_{ij}^{[\alpha]}\}$, where $a_{ij}^{[\alpha]}=1$ if node $i$ and $j$ are connected in layer $\alpha$, and $a_{ij}^{[\alpha]}=0$, otherwise. 

For each layer $\alpha$, we apply the $k$-core decomposition algorithm and assign each node $i$ a coreness value $k_i^{[\alpha]}$. Additionally, we compute its centrality value $\widetilde{m}_i^{[\alpha]}$ within the $k_i^{[\alpha]}$-core. The richness of node $i$ at layer $\alpha$ is then defined as $\boldsymbol{\mu}_i^{[\alpha]}=(k_i^{[\alpha]},\widetilde{m}_i^{[\alpha]})$. Consistent with prior definitions, when eigenvalue centrality (or degree centrality) is used, the method is referred to as MCC-E (or MCC-D). In the most simple case, we aggregate all layers evenly to obtain the multiplex richness of node $i$ \cite{supple}: 
\begin{eqnarray}\label{muliplex_rich}  
\boldsymbol{\mu}_i&=&\frac{1}{M} \sum_{\alpha=1}^M \boldsymbol{\mu}_i^{[\alpha]} \nonumber\\
&=& (\frac{1}{M} \sum_{\alpha=1}^M k_i^{[\alpha]}, \frac{1}{M}\sum_{\alpha=1}^M \widetilde{m}_i^{[\alpha]}).
\end{eqnarray}

\begin{table*}[ht]
\centering
\caption{The properties of the five multiplex networks, including the number of nodes ($V$), the number of links $E^{[i]}$ and the maximum node coreness $M_{core}^{[i]}$ in each layer $i$, as well as the size of the identified multiplex cores ($V_{core}$) and the corresponding core density coefficient $\phi$ [as defined in Eq. (\ref{mulphi})] under different methods.}

\label{tab:MDatasets}
\resizebox{\linewidth}{!}{
\begin{tabular}{c|c|ccc|ccc|cc|cc|cc}
\toprule 

\multirow{2}*{Datasets}  & \multirow{2}*{$V$}  &\multirow{2}*{$E^{[1]}$}&\multirow{2}*{$E^{[2]}$}&\multirow{2}*{$E^{[3]}$}&\multirow{2}*{$M_{core}^{[1]}$}&\multirow{2}*{$M_{core}^{[2]}$}&\multirow{2}*{$M_{core}^{[3]}$}&\multicolumn{2}{c|}{rich-core \cite{Battiston2018}}&\multicolumn{2}{c|}{MCC-E}&\multicolumn{2}{c}{MCC-D}
\\ &&&&& &&&$V_{core}$& $\phi$&$V_{core}$& $\phi$&$V_{core}$& $\phi$
\\
  \cline{1-14} Vickers-Chan-7thGraders  &29  &240	&126	 &152&13&7&8  &20  &0.8578 &11 & \textbf{1.0000}&11 &\textbf{1.0000}
\\ Lazega-Law-Firm   &71  &717 &399  &726&14&9& 14 &28 &0.6931 &12 &\textbf{0.8787} &26 &0.7200
\\ Noordin-Top   &79 &259 & 437& 200&11&17& 5&24 & 0.5434 &20 &\textbf{0.6684} &20 &\textbf{0.6684}
\\ Celegans-Connectome   &279&514  &888 &1,703&4&7&9  &48 &0.2739 &7 &\textbf{0.9523} & 8 &0.9285
\\ Sanremo-2016   &56,562&210,308  &91,658 &10,514 &32&21&6  &332 &0.0962&26 &\textbf{0.5107} &15 &0.5047
\\
 \cline{1-14}
\end{tabular}}

\end{table*}

Furthermore, in analogy to the single-layer case as described in Sec. II (with slight modifications), for each node $i$ in layer $\alpha$,  we divide its links into two groups: those toward nodes with higher richness in layer $\alpha$ (denoted as $d_i^{[\alpha]+}$) and those toward nodes with lower richness in layer $\alpha$ (denoted as $d_i^{[\alpha]-}$). It should be noted that, in this context, we simply calculate the number of neighboring nodes with higher richness for each node $i$, as studied in \cite{Battiston2018}, instead of ranking all nodes in layer $\alpha$ and counting the links connecting node $i$ to nodes ranked higher. The two approaches are equivalent when all nodes in the network have distinct richness values. However, in practical scenarios where nodes may share the same richness, discrepancies can arise. We then define the multiplex links of a node toward richer nodes as
\begin{eqnarray}\label{muliplex_d+}  
d_i^+=\frac{1}{M} \sum_{\alpha=1}^M d_i^{[\alpha]+}. 
\end{eqnarray}

Based on the above information, we first rank the nodes according to their multiplex richness $\boldsymbol{\mu}_i$, such that the node ranked first (i.e., $r_i=1$) has the highest richness, and so on. Then, we compute $d_i^+$ for each node $i$ in order and plot it as a function of the rank $r_i$. The rank corresponding to the maximum value of $d_i^+$ defines the boundary of the core-periphery structure, with nodes ranked below this value being classified as part of the multiplex core. 

To quantify the density of the multiplex core, we define the density coefficient by integrating all layers as follows
\begin{eqnarray}\label{mulphi}  
\phi=\frac{2}{V_{core}(V_{core}-1)} \sum_{\alpha=1}^M E_{core}^{[\alpha]},
\end{eqnarray}
where $V_{core}$ represents the number of nodes in the multiplex core, and $E_{core}^{[\alpha]}$ denotes the number of links among these nodes in layer $\alpha$.

We analyze five datasets with multiplex structures, detailed as follows:\\
1.~{\bf Vickers-Chan-7thGraders} \cite{Vickers1981}.
This network consists of three layers of relationships among 7th graders in Victoria, Australia: getting along, best friends, and preferred work partners.\\
2.~{\bf Lazega-Law-Firm} \cite{Lazega2001}. 
This network represents three types of relationships between partners and associates of a corporate law firm: coworkers, friendship, and advice.\\
3.~{\bf Noordin-Top} \cite{Battiston2014}.
This network comprises three layers of relationships within an international crisis group: trust, operational interactions, and communication.\\
4.~{\bf Celegans-Connectome} \cite{Chen2006}.
This network includes three types of synaptic connections in the Caenorhabditis elegans connectome: electric (“ElectrJ”), chemical monadic (“MonoSyn”), and polyadic (“PolySyn”).\\
5.~{\bf Sanremo-2016} \cite{Domenico2020}.
This network contains three layers of social relationships among Twitter users during the Sanremo Music Festival Final: retweets, mentions, and replies.

The properties of the networks, including the number of nodes, links and the maximum coreness value in each layer (all layers share the same number of nodes), are summarized in Table II. We apply our methods, MCC-E and MCC-D, to analyze these multiplex networks. For comparison, we also provide results obtained using the multiplex rich-core method \cite{Battiston2018}, where we assume each layer in a multiplex network contributes equally. Overall, our method demonstrate superior performance in identifying dense multiplex cores, with MCC-E notably achieving the highest $\phi$ values.

\section{Discussion}
In summary, we have proposed an effective method for profiling core-periphery structures in networks. Our approach builds upon the rich-core framework by first ranking the nodes in the network according to their richness. Rather than defining a node's richness solely by its degree, we characterize it as a combination of the node's coreness and its centrality within the $k$-core. Subsequently, we compute the number of links each node has to higher-ranked nodes (denoted as $d_i^+$) and examine how this quantity varies with node rank $r$. The rank corresponding to the maximum of this quantity defines the boundary of the core-periphery structure. 

The physical interpretation of this partition can be linked to the concept of a random walker, as studied in \cite{Ma2015}. Consider a network is divided into two parts following the previous algorithm: the core ($G_c$) and the periphery ($G_p$). The probability that a random walker remains within the core is given by
\begin{eqnarray}\label{alpha}  
\alpha_c=\frac{\sum_{i,j \in G_c}a_{ij}}{\sum_{i\in G_c}k_i}=2\frac{\sum_{i=1}^c d_i^+}{\sum_{i=1}^c k_i},
\end{eqnarray}
where $k_i$ is the degree of node $i$, $a_{ij}$ is the element of the adjacency matrix, and $c$ denotes the core size. As the core expands, the probability of a random walker escaping decreases, making $\alpha_c$ an increasing function of core size $c$. However, the rate of increase changes non-monotonously with $c$, reaching a maximum at the point where $\alpha_c^{''}=0$. This condition requires, to a first approximation, the second derivative of $g(c) \equiv \sum_{i=1}^c d_i^+=\int_1^c d_i^+(y)dy$ to be $0$, indicating that $d_i^+(y)$ has an extremum (maximum), which we refer to as the core-periphery boundary.

The time complexity of our method can be analyzed by each step: (i) $k$-core decomposition. The complexity is $O(|E|)$, where $|E|$ is the number of links. (ii) Computing centrality. We consider the worst case, where the core encompasses the entire network. For degree centrality, the complexity is $O(|E|)$; for eigenvector centrality, using the power iteration method, the complexity is $O(|kE|)$ ($k$ is the number of iterations). (iii) Sorting by richness. The complexity is $O(|V|log|V|)$, where $|V|$ is the number of nodes in the network. (iv) Calculating $d_i^+$. The complexity is $O(|E|)$. (v) Identifying the turning point. The complexity is $O(|V|)$. To summarize, the overall complexity is $O(|E|+|V|log|V|)$, which indicates that our method is suitable for large-scale networks.

We applied our algorithm to $12$ real-world networks and found that, compared to the traditional rich-core method, the cores identified by our approach are generally denser. Additionally, we demonstrated that there is an upper bound for the $d^+(r)$ curve, and the top-ranked nodes lying exactly on this upper bound form a clique. Remarkably, the clique identified by this straightforward way is often close to, or even identical to the maximum clique in many real-world networks. Finally, we extended the method to multiplex networks, which offer a more realistic representation of many complex systems, and demonstrated its effectiveness in identifying dense multiplex cores, particularly on five well-studied datasets. 

Our study provides a valuable framework for effectively detecting the cores (even cliques) in diverse real-world systems spanning various fields, such as biology, sociology, and transportation. It may have some important applications. For instance, in social networks, identifying densely connected groups can be crucial for enhancing or controlling information diffusion \cite{Kitsak2010,Ruan2020}. In biology networks, detecting the core regions of the brain may help to better understand several neurological diseases and contribute to the development of related treatment strategies \cite{Battiston2018}. Finally, some limitations of our study should be noted. We here mainly focus on static networks with pairwise interactions, whereas real-world networks are often dynamic \cite{Holme2012} and involving higher-order interactions \cite{Battiston2020}, which require further investigation.

\section*{Acknowledgement}
This work was supported in part by the Key Research and Development Program of Zhejiang under Grants 2022C01018 and 2024C01025, by the National Natural Science Foundation of China under Grants  U21B2001 and 62103374, by the Zhejiang Provincial Natural Science Foundation of China under Grant ZCLY24F0302.




\begin{thebibliography}{99}

\bibitem{Newman2004}
M. E. J. Newman, ``Detecting community structure in networks," Eur. Phys. J. B {\bf 38}, 321-330 (2004).

\bibitem{Newman2006}
M. E. J. Newman, ``Modularity and community structure in networks," Proc. Natl. Acad. Sci. USA {\bf 103}, 8577-8582 (2006).

\bibitem{Fortunato2022}
S. Fortunato, M. E. J. Newman, ``20 years of network community detection,", Nat. Phys. {\bf 18}, 848-850 (2022).

\bibitem{Fortunato2016}
S. Fortunato and D. Hric, ``Community detection in networks: A user guide," Phys. Rep. {\bf 659}, 1-44 (2016).

\bibitem{Shang2020}
Y. Shang, Generalized k-core percolation in networks with community structure, SIAM J. Appl. Math. {\bf 80}, 1272 (2020).

\bibitem{Rombach2017}
P. Rombach, M. A. Porter, J. H. Fowler, and P. J. Mucha, ``Core-Periphery Structure in Networks (Revisited)," SIAM Review {\bf 59}, 619 (2017).

\bibitem{Borgatti2000}
S. P. Borgatti and M. G. Everett, ``Models of core/periphery structures," Social networks {\bf 21}, 375-395 (2000).

\bibitem{Holme2005}
P. Holme, ``Core-periphery organization of complex networks," Phys. Rev. E {\bf 72}, 046111 (2005).

\bibitem{Elliott2020}
A. Elliott, A. Chiu, M. Bazzi, G. Reinert, M. Cucuringu, ``Core-periphery structure in directed networks," Proc. R. Soc. A {\bf 476}, 20190783 (2020).

\bibitem{Gallagher2021}
R. J. Gallagher, J. G. Young, B. F. Welles, ``A clarified typology of core-periphery structure in networks," Sci. Adv. {\bf 7}, eabc9800 (2021).

\bibitem{Polanco2023}
A. Polanco, M. E. J. Newman, ``Hierarchical core-periphery structure in networks," Phys. Rev. E {\bf 108}, 024311 (2023).

\bibitem{Verma2014}
T. Verma, N. A. Araújo, and H. J. Herrmann, ``Revealing the structure of the world airline network," Sci. Rep. {\bf 4}, 5638 (2014).

\bibitem{Verma2016}
T. Verma, F. Russmann, N. A. M. Araújo, J. Nagler, H. J. Herrmann, ``Emergence of core–peripheries in networks," Nat. Comm. {\bf 7}, 10441 (2016).

\bibitem{Fagiolo2010}
G. Fagiolo, J. Reyes, and S. Schiavo, ``The evolution of the world trade web: a weighted-network analysis," J. Evol. Econ. {\bf 20}, 479-514 (2010).

\bibitem{Rossa2013}
F. D. Rossa, F. Dercole, and C. Piccardi, ``Profiling core-periphery network structure by random walkers," Sci. Rep. {\bf 3}, 1467 (2013).

\bibitem{Elliott2014}
M. Elliott, B. Golub, and M. O. Jackson, ``Financial networks and contagion," Am. Econ. Rev. {\bf 104}, 3115-3153 (2014).

\bibitem{Masuda2006}
N. Masuda and N. Konno, ``Vip-club phenomenon: Emergence of elites and masterminds in social networks," PLoS One {\bf 28}, 297-309 (2006).

\bibitem{Lee2014}
S. H. Lee, M. Cucuringu, and M. A. Porter, ``Density-based and transport-based core-periphery structures in networks," Phys. Rev. E {\bf 89}, 032810 (2014).

\bibitem{Kojaku2017}
S. Kojaku and N. Masuda, ``Finding multiple core-periphery pairs in networks," Phys. Rev. E {\bf 96}, 052313 (2017).

\bibitem{Zhang2015}
X. Zhang, T. Martin, M. E. J. Newman, ``Identification of core-periphery structure in networks," Phys. Rev. E {\bf 91}, 032803 (2015).

\bibitem{Ma2015}
A. Ma and R. J. Mondragón, ``Rich-cores in networks," PLoS One {\bf 10}, e0119678 (2015).

\bibitem{Mondragon2017}
R. J. Mondragón, ``Network partition via a bound of the spectral radius," J. Complex Netw. {\bf 5}, 513-526 (2017).

\bibitem{Ma2015pnas}
A. Ma, R. J. Mondragón, and V. Latora, ``Anatomy of funded research in science," Proc. Natl Acad. Sci. USA {\bf 112}, 14760-14765 (2015).

\bibitem{Battiston2018}
F. Battiston, J. Guillon, M. Chavez, V. Latora, and F. De Vico Fallani, ``Multiplex core–periphery organization of the human connectome," J. R. Soc. Interface {\bf 15}, 20180514 (2018).

\bibitem{Kitsak2010}
M. Kitsak, L. K. Gallos, S. Havlin, F. Liljeros, L. Muchnik, H. E. Stanley, and H. A. Makse, ``Identification of influential spreaders in complex networks," Nat. Phys. {\bf 6}, 888 (2010).

\bibitem{Zachary1977}
W. W. Zachary, ``An information flow model for conflict and fission in small groups," J. Anthropol. Res. {\bf 33}, 452-473 (1977).

\bibitem{Doph2003}
D. Lusseau, K. Schneider, O. J. Boisseau, P. Haase, E. Slooten, and S. M. Dawson, ``The bottlenose dolphin community of doubtful sound features a large proportion of long-lasting associations: can geographic isolation explain this unique trait?" Behav. Ecol. Sociobiol. {\bf 54}, 396-405 (2003).

\bibitem{Knuth1993}
D. E. Knuth, The Stanford GraphBase: a platform for combinatorial computing, (AcM Press, New York, 1993).

\bibitem{Gleiser2003}
P. M. Gleiser and L. Danon, ``Community structure in jazz," Adv. Complex Syst. {\bf 6}, 565-573 (2003).

\bibitem{Viswanath2009}
B. Viswanath, A. Mislove, M. Cha, and K. P. Gummadi, ``On the evolution of user interaction in facebook," In Proceedings of the 2nd ACM workshop on Online social networks (ACM, 2009), pp. 37-42.

\bibitem{Guimera2003}
R. Guimera, L. Danon, A. Diaz-Guilera, F. Giralt, and A. Arenas, ``Self-similar community structure in a network of human interactions," Phys. Rev. E {\bf 68}, 065103 (2003).

\bibitem{newman2006finding}
M. E. Newman, ``Finding community structure in networks using the eigenvectors of matrices," Phys. Rev. E 74, 036104 (2006).


\bibitem{CiteSeer}
R. A. Rossi and N. K. Ahmed, ``The network data repository with interactive graph analytics and visualization,'' In Proceedings of 29th AAAI Conference on Artificial Intelligence (ACM, 2015), pp. 4292-4293.

\bibitem{CAIDA}
See https://catalog.caida.org/dataset/as\_relationships\_serial\_1 for information about the AS relationship network.

\bibitem{KONECT}
J. Kunegis. ``KONECT --- The Koblenz Network Collection." In Proceedings of the 22nd International Conference on World Wide Web Companion (ACM, 2013), pp. 1343-1350. 

\bibitem{Bron1973}
C. Bron and J. Kerbosch, ``Algorithm 457: finding all cliques of an undirected graph," Commun. ACM {\bf 16}, 575-577 (1973).

\bibitem{Tomita2006}
E. Tomita, A. Tanaka, and H. Takahashi, ``The worst-case time complexity for generating all maximal cliques and computational experiments," Theor. Comput. Sci. {\bf 363}, 28-42 (2006).

\bibitem{Cazals2008}
F. Cazals and C. Karande, ``A note on the problem of reporting maximal cliques," Theor. Comput. Sci. {\bf 407}, 564-568 (2008).

\bibitem{Boccaletti2014}
S. Boccaletti, G. Bianconi, R. Criado, C.I. del Genio, J. Gómez-Gardeñes, M. Romance, I. Sendiña-Nadal, Z. Wang, and M. Zanin, ``The structure and dynamics of multilayer networks," Phys. Rep. {\bf 544}, 1-122 (2014).

\bibitem{Lee2015}
K. M. Lee, B. Min, and K. I. Goh, ``Towards real-world complexity: an introduction to multiplex networks," Eur. Phys. J. B {\bf 88}, 1-20 (2015). 

\bibitem{Shen2020}
G. Shen, X. Fan, and Z. Ruan, ``Totally asymmetric simple exclusion process on multiplex networks," Chao {\bf 30}, 023103 (2020).

\bibitem{Ruan2015}
Z. Ruan, C. Wang, P. M. Hui, and Z. Liu, ``Integrated travel network model for studying epidemics: Interplay between journeys and epidemic," Sci. Rep. {\bf 5}, 11401 (2015).

\bibitem{Shang2020b}
Y. Shang, Generalized k-core percolation on correlated and uncorrelated multiplex networks, Phys. Rev. E {\bf 101}, 042306 (2020).

\bibitem{Shang2022}
Y. Shang, Feature-enriched core percolation in multiplex networks, Phys. Rev. E {\bf 106}, 054314 (2022).

\bibitem{supple}
In some cases, different layers may differ significantly in link density. To address this imbalance, one approach is to weight richness by the maximum coreness and maximum centrality value in each layer. 

\bibitem{Vickers1981}
M. Vickers and S. Chan, ``Representing classroom social structure," Victoria Institute of Secondary Education, Melbourne (1981).

\bibitem{Lazega2001}
E. Lazega, The collegial phenomenon: The social mechanisms of cooperation among peers in a corporate law partnership (Oxford University Press, USA, 2001).

\bibitem{Battiston2014}
F. Battiston, V. Nicosia, V Latora, ``Structural measures for multiplex networks," Phys. Rev. E {\bf 89}, 032804 (2014).

\bibitem{Chen2006}
B. L. Chen, D. H. Hall, and D. B. Chklovskii, ``Wiring optimization can relate neuronal structure and function," Proc. Natl Acad. Sci. USA {\bf 103}, 4723-4728 (2006).

\bibitem{Domenico2020}
M. De Domenico and E. G. Altmann, ``Unraveling the origin of social bursts in collective attention," Sci. Rep. {\bf 10}, 4629 (2020).




\bibitem{Ruan2020}
Z. Ruan, B. Yu, X. Shu, Q. Zhang, and Q. Xuan, The impact of malicious nodes on the spreading of false information, Chaos {\bf 30}, 083101 (2020).

\bibitem{Holme2012}
P. Holme and J. Saramaki, Temporal networks, Phys. Rep. {\bf 519}, 97 (2012).


\bibitem{Battiston2020}
F. Battiston, G. Cencetti, I. Iacopini, V. Latora, M. Lucas, A. Patania, J. G. Young, and G. Petri, Networks beyond pairwise interactions: structure and dynamics, Phys. Rep. {\bf 874}, 1 (2020).


\end{thebibliography}
\end{document}